\documentstyle[preprint,aps,epsf]{revtex}                  %%%To be commented
\global\let\epsfloaded=Y %\usepackage{epsfig}
\begin{document}
\pagestyle{empty}                                      %%%To be commented
\preprint{
\font\fortssbx=cmssbx10 scaled \magstep2
%\hbox to \hsize{
%\hbox{%\hskip.2cm
%            hep-ph/9804xyz}
%\hfill $%\raise .5cm
%\vtop{              
%                %\hbox{NTUTH-97-05}
% \hbox{ }}$
%}
}
\draft
\vfill
\title{
$CP$ Violating Phase Difference between $B\to J/\psi K_S$ and  $J/\psi K_S \pi^0$
from New Physics
}

\vfill
\author{$^{1,2}$Xiao-Gang He and $^2$Wei-Shu Hou}
\address{
\rm $^1$ School of Physics, University of Melbourne, 
Parkville, Vic. 3052, Australia\\
\rm $^2$Department of Physics, National Taiwan University,
Taipei, Taiwan 10764, R.O.C.
}

%\date{\today}
%
%\vskip -1cm
%
\vfill
\maketitle
\begin{abstract}
The small P wave compoment of $B\rightarrow J/\psi K^*$ 
measured by CLEO makes it practical to measure $\sin2\beta$ 
using $B^0\rightarrow J/\psi K^{*0}\rightarrow J/\psi K_S \pi^0$,
independent from $B^0\rightarrow J/\psi K_S$.
Because these modes are color suppressed,
new physics enhanced color dipole $bsg$ coupling, 
as hinted from the persistent ${\cal B}_{s.l.}$ and $n_C$ problems as well as
the newly observed large semi-inclusive $B\to \eta^\prime + X_s$ decay, 
may have significant impact.
We show that it may lead to a difference in the $\sin 2\beta$ measurements
between $B^0\to J/\psi K_S$ and  $J/\psi K_S \pi^0$ measureable
at the B Factories in the near future.
\end{abstract}
%
%\vfill
%
%
%\narrowtext
%
\pagestyle{plain}

%\section{Introduction}

The CLEO Collaboration has recently reported \cite{Pwave} 
the first full angular analysis of the color-suppressed  decays 
$B\to  J/\psi K^{*0}$ and  $J/\psi K^{*+}$.
They find that the $P$ wave component is small,
$\vert P\vert^2 = 0.16 \pm 0.04$,
which means that $B^0\to  J/\psi K^{*0} \to J/\psi K_S \pi^0$ decay is
dominated by $CP$-even  final states.
The mode can therefore be used to measure 
the $CP$ violating angle $\sin 2\beta$
without the complication of an angular analysis \cite{DQSTL}.
Compared to the gold-plated ($CP$-odd) $J/\psi K_S$ mode,
one has a dilution factor $\sim 30\%$ but this is now already measured,
hence both modes can be profitably studied as the B Factories turn on in 1999.
An interesting question can now be raised:
What if  $\sin 2\beta_{J/\psi K_S} \neq  \sin 2\beta_{J/\psi K_S \pi^0}$?
Naively it is hard to conceive how
$\sin 2\beta_{J/\psi K_S}$ and $\sin 2\beta_{J/\psi K_S \pi^0}$ could differ,
since the decays are from the {\it tree level} CKM dominant $b\to c\bar cs$ process,
while any change in $CP$ violating phase due to new physics
in $B^0$--$\bar B^0$ mixing is just a common factor.
One needs new contribution to the decay amplitudes.
In this paper we elucidate some new physics mechanisms whereby 
the $\sin 2\beta$ measurements could differ between  
$B \to J/\psi K_S$ and $B\to  J/\psi K_S \pi^0$ decays,
hence illustrating the importance of making and refining
these separate measurements.

Let us first illustrate how 
$\sin2\beta_{J/\psi K_S}$ and $\sin 2 \beta_{J/\psi K_S \pi^0}$ could in principle differ.
The mixing-dependent CP violation measurable 
is ${\rm Im\,}\xi = {\rm Im\,}\{(q/p)(A^*\bar A/\vert A\vert ^2)\}$,
where $q/p = e^{-2i\phi_B}$ is from $B^0$--$\bar B^0$ mixing, 
$A=\vert A\vert e^{i\phi_w}$ and $\bar A= \vert \bar A\vert  e^{-i\phi_w}$ are 
decay amplitudes for $B$ and  $\bar B$ decays into the same CP eigenstates, 
respectively. 
For $B\rightarrow J/\psi K_S$,
the final state is purely CP odd. Taking the weak phase of 
the decay amplitude to be $\phi_0$, one has
\begin{eqnarray}
{\rm Im\,}\xi(B\rightarrow J/\psi K_S)= - \sin2\beta_{J/\psi K_S} = 
-\sin(2\phi_B + 2 \phi_0).
\end{eqnarray}
For $B\rightarrow J/\psi K^* \rightarrow J/\psi K_S \pi^0$, the final state
is a  mixture of CP odd and even states. 
Taking $\phi_w$ for these decay amplitudes 
to be $\phi_1$ and $\tilde \phi_1$, respectively,
one obtains
\begin{eqnarray}
{\rm Im\,}\xi(B\rightarrow J/\psi K_S \pi^0)
&=& {\rm Im\,} \{ e^{-2i\phi_B} [ e^{-2i\phi_1}\vert P \vert ^2
                                                      - e^{-2i\tilde \phi_1} (1-\vert P \vert ^2)]\}\nonumber\\
&\equiv& (1-2\vert P \vert ^2)\sin2\beta_{J/\psi K_S\pi^0},
\end{eqnarray}
where $\vert P \vert ^2$ is the CP odd fraction.
We have defined $\sin2\beta_{J/\psi K_S\pi^0}$ in such a way that 
in the Standard Model (SM) it is equal to $\sin2\beta$,
that is %One then obtains the usual result of
\begin{eqnarray}
\ \ \ \ \ \ \ \ \ \ \ \ \ \ \ \ 
\sin2\beta_{J/\psi K_S}= \sin2\beta_{J/\psi K_S\pi^0}= \sin2\beta,
 \ \ \ \ \ \ \ {\rm (Standard\ Model)}
\end{eqnarray}
since one has $\phi_0 = \phi_1 = \tilde \phi_1$ and $\phi_B+\phi_0 = \beta$.
In the Wolfenstein  parametrization, which we adopt, 
one has $\phi_0 = 0$ and $\phi_B = \beta$.
Clearly, both measurements provide true information about $\sin2\beta$
within SM.
However, this is no longer true if one goes beyond SM. 
The two phases $\phi_1$ and $\tilde \phi_1$
may differ, and they may also be different from $\phi_0$. 
If such is the case, then
$\sin2\beta_{J/\psi K_S} \neq \sin2\beta_{J/\psi K_S\pi^0}$ follows.

There are many ways where new physics may change 
the phases $\phi_{0,1}$ and $\tilde \phi_1$. 
To lowest order they are through
 dimension 6 four quark operators, or else the
dimension 5 color dipole operator $\bar s i\sigma_{\mu\nu}G^{\mu\nu} 
(1\pm \gamma_5) b$, where $G^{\mu\nu}$ is the gluon field strength.
New physics contributions of the
form $\bar c \gamma^\mu(1\pm \gamma_5) c \bar s \gamma_\mu (1-\gamma_5) b$
may generate a common phase shift for $\phi_{0,1}$ and $\tilde \phi_1$, 
whereas operators of the form
$\bar c\gamma^\mu (1\pm \gamma_5) c \bar s \gamma_\mu (1+\gamma_5) b$
may shift $\phi_0$ and $\phi_1$ by a common factor and 
$\tilde \phi_1$ by the same amount but with opposite sign. 
Our primary example, however, would be 
the dimension 5 color dipole operator,
since there are experimental hints that 
it may be large in Nature. 

We parametrize the color dipole interaction as
\begin{eqnarray}
 {G_F\over \sqrt{2}} V_{tb}V_{ts}^* {g_s\over 16\pi^2} 
F_2 m_b\, \bar s \sigma_{\mu\nu} G^{\mu\nu} (1+\gamma_5)b.
\end{eqnarray}
In SM $F_2 \simeq 0.286$ is small, and the process $b\to sg$ 
(where $g$ is ``on-shell", or jet-like) is only of order 0.2\%, 
and is very hard to measure experimentally.
However, the persistent low semileptonic $B$ branching ratio
(${\cal B}_{s.l.}$) and charm counting ($n_C$) problems suggest 
that $b\to sg$ decay could be enhanced to the $10\%$ level \cite{bsg,Kagan},
which implies  $\vert F_2 \vert \sim 2$.
A bound on $b\to sg$ from the recent $B\to D\bar DK+X$ study \cite{UpperVertex}
does not yet rule this out, while the discovery of a surprisingly large
semi-inclusive $B\to \eta^\prime + X_s$ decay \cite{etapXs}
may call for an interplay of \cite{HT} 
the SM $bsg$ charge radius coupling $\vert F_1^{\rm SM} \vert \sim 5$ 
and the new physics dipole coupling $\vert F_2 \vert \sim 2$.
Furthermore, this new physics enhanced dipole coupling brings in
naturally a CKM-indendent $CP$ violating phase, 
and could lead to rate asymmetries at the 10\% level \cite{HT}
in the $m_{X_s}$ recoil mass spectrum of the $B\to \eta^\prime + X_s$ mode.

Still, how can $\phi_{0,1}$ and $\tilde \phi_1$ be 
significantly changed by the color dipole interaction?
Note that  $B\to J/\psi K_S(K^*)$ decays are color suppressed.
Assuming factorization, the decay rate is 
proportional to $\vert c_1 + c_2/N\vert^2$,
which suffer from accidental cancellation for $N = 3$.
A phenomenological fit suggests $N_{\rm eff.} \simeq 2$ \cite{Neff,Neubert}, or,
alternatively, if one takes $N= 3$ from QCD,
then there must be sizable color-octet and other nonfactorizable contributions.
The former amplitude is found to be $\sim 1/N$,
similar to the color-singlet term, while 
the other nonfactorizable contributions are $\sim 1/N^2$ \cite{Neubert} 
in the large $N$ expansion.
%It has been shown that the inclusion of color-octet contributions
%can significantly 
%improve \cite{KLS,HeSoni} the agreement with data for $N = 3$.
Thus, the weak phases $\phi_{0,1}$ and $\tilde \phi_1$ should be 
sensitive to color octet operators such as the color dipole.

Even for  $b\to sg \sim 10\%$, by itself its contribution to 
$B\to J/\psi K^*$ is only a small fraction of the measured $\sim 10^{-3}$ rate.
However, the shift in $2\beta$ for $J/\psi K_S \pi^0$ mode could
be at $-0.15\sin\phi$ level, where $\phi$ is the new $CP$ violating phase.
In contrast, because of the dipole nature of the $F_2$ coupling,
the corresponding shift in the $B\to J/\psi K_S$ mode is suppressed by a factor 
$m_\psi^2/m_B^2$.
The phase difference is clearly measurable in the near future.

We now turn to some details.
In the factorization approximation but including color octet contributions, 
the leading order decay amplitudes for $B\rightarrow J/\psi K_S(K^*)$ 
can be written as  
\begin{eqnarray}
A (B \rightarrow J/\psi K_S(K^*))
&=& i {G_F \over \sqrt{2}} V_{cb}V_{cs}^* f_\psi m_\psi  \varepsilon_\psi ^\mu
\left\{ (B_1 +2B_8 r_8)
\langle K_S(K^*)\vert \bar s \gamma_\mu (1-\gamma_5) b\vert B\rangle
\phantom{{\alpha_s \over m^2_\psi}} \right .\nonumber\\
&+&\left . {\alpha_s \over 2 \pi} {m_b\over q^2} r_8^\prime \langle K_S(K^*)\vert 
\bar s i \sigma_{\mu\nu} q^\nu (c_8(1+\gamma_5) + \tilde c_8 (1-\gamma_5))
b\vert B\rangle
\right \},
\end{eqnarray}
where we have adopted $c_8 = -F_2$ in the operator language,
and
\begin{eqnarray}
&&B_1 = c_1+{c_2\over N} - \sum_{i=u,c,t}
 {V_{ib}V_{is}^*\over V_{cb}V_{cs}^*}
\left(c^i_3+{c_4^i\over N} + c_5^i+{c_6^i\over N}\right),\nonumber\\
&&B_8 = c_2 - \sum_{i=u,c,t} {V_{ib}V_{is}^*\over V_{cb}V_{cs}^*}
\left (c_4^i+c_6^i\right),
\end{eqnarray}
where the Wilson coefficients $c_j^i$ for tree and strong penguin operators 
evaluated in Ref.\cite{desh-he} will be used. 
The index $i$ indicates the quark in the penguin loop. 
The electroweak penguin contributions have been neglected. 
The decay constant $f_\psi$ is defined by
$\langle J/\psi \vert  \bar c \gamma^\mu c\vert 0\rangle 
= i f_\psi m_\psi \varepsilon^\mu_\psi$, 
and is determined from leptonic decays of $J/\psi$ to be 
$f_\psi \simeq 410$ MeV.
The parameters $r_8$ and $r_8^\prime$ are ratios of color octet 
and singlet matrix elements
\begin{eqnarray}
r_8 &=&{ \langle J/\psi K_S(K^*)\vert  [\bar c \gamma_\mu T^a c] \, 
[\bar s\gamma^\mu (1-\gamma_5)T^a b] \vert B\rangle
 \over
\langle J/\psi \vert \bar c \gamma^\mu c\vert 0\rangle
\langle K_S(K^*)\vert \bar s \gamma^\mu (1-\gamma_5) b\vert B\rangle},
\nonumber\\
r_8^\prime&=& {\langle J/\psi K_S(K^*) \vert  [\bar c \gamma^\mu T^a c] \, 
[\bar s i\sigma_{\mu\nu}q^\nu (1\pm \gamma_5)T^a b]\vert B\rangle
 \over 
\langle J/\psi \vert \bar c \gamma^\mu c\vert 0\rangle \langle K_S(K^*)\vert 
\bar s i\sigma_{\mu\nu} q^\nu (1\pm \gamma_5)b\vert B\rangle}.
\end{eqnarray}
These two parameters can in principle be different, 
but at the moment it is not possible to calculate them from first principles. 
We will determine $r_8$ by fitting experimental data, 
and take $r_8^\prime = r_8$ to be equal.
This will give an indication of the size of the effect. 
We also include a possible color dipole 
$(1-\gamma_5)$ term with strength $\tilde c_8$ for completeness.

Following the notation of Ref.\cite{bsw}, we parametrize
\begin{eqnarray}
\langle \bar K^0\vert  \bar s \gamma_\mu (1-\gamma_5) b \vert  B\rangle 
= F_1(q^2) \left(p^B_\mu + p^K_\mu\right) 
+ {m_B^2-m_K^2\over q^2} \left(F_0(q^2) - F_1(q^2)\right) q_\mu,
\end{eqnarray}
where we take $F_1(0) = F_0(0) = 0.379$.
For the $q^2$ dependence of these form factors, we assume 
the pole form $F_i(q^2) = F_i(0)/(1-q^2/m_i^2)^n$, 
with $m_0 = 5.98$ GeV and $m_1 = 5.43$ GeV. 
Heavy quark effective theory suggests that at maximum recoil, 
$F_1(q^2_{\rm max})/F_0(q^2_{\rm max})$ scales as $m_b$\cite{iw},
which implies that if the pole power for $F_1(q^2)$ is $n$, 
then $F_0(q^2)$ is $n-1$. 
We take dipole behavior for $F_1$ and monopole for $F_0$.

At maximum recoil $\langle \bar K^0\vert \bar s i\sigma_{\mu\nu} q^\nu 
(1\pm \gamma_5) b \vert B\rangle$ can be related to 
the form factors defined in Eq. (8) by heavy quark effective theory\cite{iw}. 
With $m_b \simeq m_B$, $p_b \simeq p_B$ and 
extrapolating to $q^2 \cong m_{J/\psi}^2$, we find
\begin{eqnarray}
\varepsilon_\psi^\mu \, \langle \bar K^0\vert  
\bar s i\sigma_{\mu\nu} q^\nu (1\pm \gamma_5) b \vert B\rangle
\cong 
2 \varepsilon_\psi \cdot p_B\,m_B\,F_1(q^2) \, s(q^2),
\end{eqnarray}
where $s(q^2)= \left[(F_0(q^2)/F_1(q^2)-1)(1-m_K^2/ m_B^2) - q^2/ m_B^2\right]/2
 \approx -m^2_\psi/m_B^2$ is a suppression factor due to the helicity structure of
$\sigma_{\mu\nu}$ at the quark level.
Note that in the photon case one has $q^2 = 0$ and this factor vanishes, as it should.
The full decay amplitude is given by
\begin{eqnarray}
A(B\rightarrow J/\psi K_S) &\cong & iG_F V_{cb}V_{cs}^* f_\psi m_\psi 
F_1(m_\psi^2)\, \varepsilon_\psi \cdot p_B \nonumber\\
& \times &  \left\{ \left[ B_1 + 2 B_8 r_8 \right] + {\alpha_s \over 2\pi} {m_B^2\over m_\psi^2} \left[ c_8+ \tilde c_8 \right] r_8 \, s(m_\psi^2) \right \}.
\end{eqnarray}
To fit experimental data, one needs 
$1/N_{eff}=1/N +2r_8  \approx 1/2$ \cite{Neff,Neubert}, giving
$r_8 \approx 1/12$.  

For $B\rightarrow J/\psi K^*$, 
we continue to use the form factor parametrization of \cite{bsw},
\begin{eqnarray}
\langle K^* \vert  \bar s \gamma_\mu (1-\gamma_5) b\vert B\rangle
&=& {2V(q^2)\over m_B + m_{K^*}} \, \varepsilon_{\mu\nu\alpha\beta} \, \varepsilon_{K^*}^\nu q^\alpha p_{K^*}^\beta
+ i {A_2(q^2)\over m_B+m_{K^*}} \,\varepsilon_{K^*}\cdot q\, (p_B^\mu + p_{K^*}^\mu)
\nonumber\\
&-&  i(m_B+m_{K^*}) A_1(q^2) \, \varepsilon_{K^*}^\mu 
+2im_{K^*}{A_3(q^2)-A_0(q^2)\over q^2} \,\varepsilon_{K^*}\cdot q \,q_\mu,
\end{eqnarray}
where $2m_{K^*}A_3(q^2) = (m_B+m_{K^*})A_1(q^2)- (m_B-m_{K^*})A_2(q^2)$,
and $V(0),\ A_1(0),\ A_2(0) = 0.369, \ 0.328, \ 0.331$ 
with pole masses $5.43$,  $5.98$ and $5.82$ GeV, respectively,
while  $A_0(0) = A_3(0)$ with pole mass 5.38 GeV. 
We assume dipole behavior for $V(q^2),\;A_2(q^2)$ and monopole for
$A_1(q^2),\;A_0(q^2)$.
For $\langle K^*\vert \bar s i\sigma_{\mu\nu} b \vert B \rangle$, 
we parametrize 
\begin{eqnarray}
\langle K^*\vert \bar s i\sigma_{\mu\nu} b \vert B \rangle 
&=& \varepsilon_{\mu\nu\alpha\beta} \left(
          g_+(q^2) \, \varepsilon_{K^*}^\alpha (p_B+p_{K^*})^\beta
  +     g_-(q^2) \, \varepsilon_{K^*}^\alpha (p_B-p_{K^*})^\beta \right.\nonumber\\
&& \ \ \ \ \ \, + \left.h(q^2) \, 
            (p_B+p_{K^*})^\alpha (p_B-p_{K^*})^\beta \, \varepsilon_{K^*}\cdot p_B\right).
\end{eqnarray}
The matrix element 
$\langle K^*\vert \bar s i\sigma_{\mu\nu}\gamma_5 b \vert B \rangle$ 
is related to the above by the identity 
$i\sigma^{\mu\nu} = (1/2) \epsilon^{\mu\nu\alpha\beta} 
\sigma_{\alpha\beta} \gamma_5$ where $\epsilon^{0123} = 1$. 
Using heavy quark effective theory at maximum recoil and 
extrapolating down to the desired $q^2$, we have
\begin{eqnarray}
g_+(q^2) &=& {1\over 2}
\left [ {m_B+m_{K^*}\over m_B} A_1(q^2)
+ {m_B^2-m_{K^*}^2+q^2\over m_B(m_B + m_{K^*})} V(q^2) \right ],
\,\nonumber\\
g_-(q^2) &=& {1\over 2}
\left [ {m_B+m_{K^*}\over m_B} A_1(q^2)
- {3m_B^2+m_{K^*}^2-q^2\over m_B(m_B + m_{K^*})} V(q^2)\right ],
\,\nonumber\\
h(q^2) \ &=&
\phantom{1\over 2} \left [ {V(q^2)\over m_B(m_B+m_{K^*})}
-{A_2(q^2)\over 2 m_B(m_B+m_{K^*})}
-{m_{K^*}A_0(q^2)\over m_B q^2}\right .\,\nonumber\\
&& + \left . {1\over 2m_B q^2}
\left ((m_B+m_{K^*})A_1(q^2) - (m_B-m_{K^*})A_2(q^2)\right)\right ].
\end{eqnarray}

The full decay amplitude is given by
\begin{eqnarray}
A(B&\rightarrow& J/\psi K^*) = i\sqrt{2}G_FV_{cb}V_{cs}^* f_\psi m_\psi 
\left\{ \left[  B_1 +2B_8 r_8 \right] \left( {V(m_\psi^2)\over m_B+m_{K^*}} \, 
\varepsilon_{\mu\nu\alpha\beta} \, \varepsilon_\psi^\mu \varepsilon_{K^*}^\nu 
p_B^\alpha p_{K^*}^\beta \right. \right.\nonumber \\
& -& \left. {i\over 2}(m_B+m_{K^*}) A_1(m_\psi^2) \, \varepsilon_\psi \cdot \varepsilon_{K^*}
+ i{A_2(m_\psi^2)\over m_B+m_{K^*}}\, 
\varepsilon_\psi\cdot p_B \, \varepsilon_{K^*}\cdot p_B\right)  \nonumber\\
&-&\ {\alpha_s\over 2\pi} {m_B\over m_\psi^2} \left[ c_8+\tilde c_8 \right] r_8
\, g_+(m_\psi^2) \, \varepsilon_{\mu\nu\alpha\beta} \, \varepsilon_\psi^\mu \varepsilon_{K^*}^\nu 
p_B^\alpha p_{K^*}^\beta\,\nonumber\\
&+& i{\alpha_s\over 2\pi} {m_B\over m_\psi^2} \left[ c_8-\tilde c_8 \right]  r_8 
\left( {1\over 2} 
\left(g_+(m^2_\psi)(m_B^2-m_{K^*}^2) +g_-(m^2_\psi) m_\psi^2\right) 
\varepsilon_\psi\cdot \varepsilon_{K^*} \phantom{\vert \over 2}\right .\nonumber\\
&& \left .\left .  \phantom{V(m_\psi^2)\over m_B+m_{K^*}V(M)} - \left( g_+(m^2_\psi)-h(m_\psi^2)m^2_\psi \right)
\varepsilon_\psi\cdot p_B \varepsilon_{K^*}\cdot p_B \right) \right \}.
\end{eqnarray}

Unlike the $s(m^2_\psi)$ factor in Eq. (10), 
the form factors $g_+(m^2_\psi)$ and $h(m_\psi^2)m^2_\psi$ in Eq. (14) 
are not suppressed by $m^2_\psi/m_B^2$.
Thus, the color dipole contribution to $B\rightarrow J/\psi K_S$
is suppressed by a factor of $m_\psi^2/m_B^2$ compared to $B\rightarrow J/\psi K^*$.
This suppression factor can be traced to the helicity structure of the
$\sigma_{\mu\nu}$ interaction at the quark level. 
Although we have worked with color singlet operators, 
we note that QCD is helicity conserving.
Soft  gluon emissions would not change the helicity structure, 
and hence the color octet dipole contribution to $B\rightarrow J/\psi K_S$ 
should still be suppressed compared to $B\rightarrow J/\psi K^*$.
If the color dipole operator coefficients $c_8$ or $\tilde c_8$ contain CP violating phases, 
the measurements of $\sin2\beta_{J/\psi K_S}$ and $\sin2\beta_{J/\psi K_S\pi^0}$
could differ.

The source of CP violation must come from physics beyond SM. 
We consider one possibility here.
It has been shown that in supersymmetric models, it is possible 
to have large $c_8$ in the desired range ($\vert c_8 \vert \approx 2$) 
necessary for solving the $n_c$ and ${\cal B}_{s.l.}$ problems,
from exchange of gluino and squarks in the loop \cite{Kagan,cuichini}.
The SUSY contribution to $c_8$ is given by %\cite{cuichini}
\begin{eqnarray}
c_8 &=& {\sqrt{2} \pi \alpha_s \over V_{tb}V_{ts}^* G_F}
\left \{(U_{LL}^\dagger \tilde M^2_{L}U_{LL})_{sb}
       {g_3(m^2_{\tilde g}/\tilde m^2_{L})\over \tilde m^4_{L}}
+ (U_{RL}^\dagger \tilde M^2_R U_{RL})_{sb}
       {g_3(m^2_{\tilde g}/\tilde m^2_{R})\over \tilde m^4_{R}}
\right .\nonumber\\
&+& \left .(U_{RL}^\dagger \tilde M^2_R U_{RR})_{sb}
       {g_4(m^2_{\tilde g}/\tilde m^2_{R})\over m_b \tilde m^3_{R}}
+(U_{LL}^\dagger \tilde M^2_LU_{LR})_{sb}
       {g_4(m^2_{\tilde g}/\tilde m^2_{L})\over m_b \tilde m^3_{L}}\right \},
\end{eqnarray}
where $m^2_{\tilde g}$ is the gluino mass, 
$\tilde M^2_{L,R}$ are the diagonal down squark mass matrices, 
and we have used the average squark masses $\tilde m^2_{L,R}$ in the
functions $g_{3,4}$, which are given in Ref.\cite{cuichini}.
The $U$ matrices transform weak basis $(\tilde D_R^0, \tilde D_L^0)$
 to mass basis $(\tilde D_R, \tilde D_L)$ squarks,
\begin{eqnarray}
\left ( \begin{array}{l}
\tilde D_R\\
\tilde D_L
\end{array}
\right )
=
\left (\begin{array}{ll}
U_{RR} & U_{RL}\\
U_{LR} & U_{LL} 
\end{array}
\right )
\left ( \begin{array}{l}
\tilde D_R^0\\
\tilde D_L^0
\end{array}
\right ).
\end{eqnarray}
One obtains $\tilde c_8$ by exchanging $L$ and $R$ in Eq. (15).
There will also be corrections to $F_1$ (or $c_{3-6}$), 
but these are subleading compared to the large log 
already contained in the SM contribution,
since SUSY corrections are from heavy internal particles 
with masses at same order of magnitude \cite{abel}.
In any case, inclusion of $F_1$ corrections will not change our conclusion,
and in principle one can find some parameter 
space in which the corrections are much smaller than the SM one. 

It is evident that $c_8$ of Eq. (15) contains in general  many phases.
For illustration let us consider the special case where
only the second and third generation squarks mix and there are
no mixings between $\tilde D_R$ and $\tilde D_L$. 
Then
\begin{eqnarray}
c_8(\tilde c_8) = {\sqrt{2}\pi \alpha_s\over V_{tb}V_{ts}^* G_F} \tilde m^2_{23} e^{i\phi}
g_3(m^2_{\tilde g}/\tilde m^2_{L(R)})\;.
\end{eqnarray}
where $\tilde m^2_{23}e^{-i\phi}$ is the 2-3 entry in the squark mass matrices. 
This scenario is least constrained by experiment. 
In particular, the phase $\phi$ is not constrained by 
low energy phenomena involving the first generation,
and the $B^0-\bar B^0$ mixing phase $\phi_B$ is also unaffected.
In the following we use this model to illustrate the shift in $\sin2\beta$. 

As has been discussed earlier,  the unique feature of color dipole coupling
is that its effect on $B\rightarrow J/\psi K_S$ is small, whereas 
the impact on $B\rightarrow J/\psi K^*$ is enhanced by $m_B^2/m_\psi^2 \approx 3$. 
The numerical value depends on $\alpha_s(m_b)$, which we will vary from 0.2 to 0.25. 
For $| c_8|=2$ and $\tilde c_8=0$, we find that the color dipole 
contribution to the decay amplitude can be as large as 8\%. 
If $\phi \neq 0$, 
it would generate phases for the CP even and odd amplitudes
\begin{eqnarray}
\phi_0 &\approx& (0.02-0.03) \sin\phi,\nonumber\\
\tilde \phi_1&\approx&\phi_1 \approx (0.06-0.08) \sin\phi,
\end{eqnarray}
becoming larger as $\alpha_s$ increases.
One would measure $\sin2\beta_{J/\psi K_s\pi^0}
\approx \sin(2\beta+0.12 \sin\phi)$ and $\sin(2\beta+0.16\sin\phi)$ for 
$\alpha_s(m_b)$ equal to $0.20$ and $0.25$ respectively.
Taking the present best fit value of 0.68 for $\sin2\beta$\cite{stocchi}, 
the enhanced color dipole interaction could 
shift $\sin2\beta_{J/\psi K_s\pi^0}$ by as much as $17\%$. 
The shift in $\sin2\beta_{J/\psi K_S}$ is three times smaller. 
Deviations between $\sin2\beta_{J/\psi K_S}$ and $\sin2\beta_{J/\psi K_S\pi^0}$ 
as large as discussed here will be probed soon at asymmetric 
B factories, and in the future at the LHC,
where $\sin2\beta_{J/\psi K_S(\pi^0)}$ can be measured to 1\% accuracy.
We note that, if $b\to sg \sim$ 10\% (hence $\vert c_8\vert \approx 2$)
sounds extreme,  $b\to sg$ at 3\% and 1\% level cannot be easily ruled out by 
experimental methods given in Ref. \cite{UpperVertex}.
Although one would then be decoupled from the $n_c$ and ${\cal B}_{s.l.}$ problems,
the difference in  $\sin2\beta_{J/\psi K_S}$ and $\sin2\beta_{J/\psi K_S\pi^0}$
would still be at $\sim 60\%$ and 30\% of those discussed here,
and still measurable.

If it turns out that $c_8$ is small but $|\tilde c_8| \sim 2$, 
$\phi_{0,1}$ remains the same, but $\tilde \phi_1$ changes sign.
We now have 
$\sin2\beta_{J/\psi K_S \pi^0} \approx sin(2\beta - 2\tilde \phi_1/(1-2\vert P\vert^2))$
and the shift is larger by a factor of $1/(1-2\vert P\vert^2)$ compared with the previous case.
The relative sign of $\phi_1 $ and $\tilde \phi_1$ can be probed by performing an
angular analysis \cite{DQSTL}.

The effects discussed here is really some form of ``penguin pollution"
due to new physics. As such, the estimates are not fully precise.
We have relied on the color octet mechanism 
to estimate the color dipole contribution.
This point needs further clarification. 
However, we find support from inclusive $B\rightarrow J/\psi X_s$ 
using the formalism of Ref. \cite{soni}. 
%, and fitting color octet contribution to experimental data.
The tree level color octet contribution 
improves agreement with inclusive rate for $N = 3$.
Through interference, 
the  color dipole operator contributes about 20\% to the decay rate, 
which is consistent with the level found in exclusive decays. 
Hence, we feel that the numerical values obtained here are 
of the right order of magnitude.

The difference between $\sin2\beta_{J/\psi K_S}$ and $\sin2\beta_{J/\psi K_S\pi^0}$
is sensitive also to other forms of new physics.
Let us give an example of possible contributions from dimension 6 operators.
In R-parity violating supersymmetric models,
exchange of charged sleptons and down type squarks
can generate currents involving the right handed quark $b_R$
with a new CP violating phase $\phi_{R}$, and therefore 
phase shifts of the type $\delta\phi_0 = \delta \phi_1 = -\delta \tilde \phi$ 
as mentioned earlier. 
The couplings involved are constrained by $b\rightarrow s \gamma$, 
but contributions at 10\% level to the amplitude is still allowed \cite{susy}. 
The difference between $\sin2\beta_{J/\psi K_S}$ and $\sin2\beta_{J/\psi K_S\pi^0}$
can be of order $0.1\sin\phi_{R} \cos \phi_B$, where we 
have used $\phi_B$ instead of $\beta$ because new physics may also 
contribute to $\phi_B$, which is in general different from $\beta$. 
The effect is again measureable. 

Before closing we would like to point out that the color dipole effect
on CP violation can be further studied in $B\rightarrow \phi K_S$ decay, where
${\rm Im\,}\xi(B\rightarrow \phi K_S)$ is also a measure of $\sin2\beta$ in SM. 
At leading order this is a pure penguin process and 
already at loop level, hence the enhanced color dipole interaction will have
larger effect than in $B\rightarrow J/\psi K^*$. 
The shift in $\sin2\beta_{\phi K_S}$ will be much larger than the case for
$B\rightarrow J/\psi K_S\pi^0$. 
A large color dipole interaction can also induce large rate asymmetries in
penguin dominated B decays such as $ B^-\rightarrow \phi K^-$ 
and $K^-\pi^0$ as well as $B^0\to K^+\pi^-$.
These modes are self-tagging hence easier to measure.
Details will be discussed elsewhere.

In conclusion, we have demonstrated that physics beyond the Standard Model
can change the prediction for $\sin2\beta_{J/\psi K_S}$ and $\sin2\beta_{J/\psi K_S\pi^0}$.
The shift due to large color dipole interaction is small for 
$\sin2\beta_{J/\psi K_S}$, but can be as large as 17\% for $\sin2\beta_{J/\psi K_S\pi^0}$.
This will be tested at B factories. The shift can be much larger for
$\sin2\beta_{\phi K_S}$.

This work is supported in part by 
grant
NSC 87-2112-M-002-037 and NSC 87-2811-M-002-046 
of the Republic of China and by Australian Research Council.

\end{document}